\documentclass[twocolumn,amsmath,preprintnumbers,superscriptaddress,amssymb]{revtex4-1}
\usepackage{graphicx}
\usepackage{xcolor}
\usepackage[T2A]{fontenc}
\usepackage{natbib}

\begin{document}
\title{Optimisation challenge for superconducting adiabatic neural network implementing XOR and OR boolean functions}

\author{D. S. Pashin}
\affiliation{Faculty of Physics, Lobachevsky State University of Nizhni Novgorod, 603022 Nizhny Novgorod, Russia}

\author{M. V. Bastrakova}
\affiliation{Faculty of Physics, Lobachevsky State University of Nizhni Novgorod, 603022 Nizhny Novgorod, Russia}
\affiliation{Russian Quantum Center, 143025 Skolkovo, Moscow, Russia}

\author{D. A. Rybin}
\affiliation{Faculty of Physics, Lobachevsky State University of Nizhni Novgorod, 603022 Nizhny Novgorod, Russia}

\author{I. I. Soloviev}
\email{igor.soloviev@gmail.com}
\affiliation{Russian Quantum Center, 143025 Skolkovo, Moscow, Russia}
\affiliation{Skobeltsyn Institute of Nuclear Physics, Lomonosov Moscow State University, 119991 Moscow, Russia, Russia}
\affiliation{National University of Science and Technology MISIS, 119049 Moscow, Russia}

\author{A. E. Schegolev}
\affiliation{Faculty of Physics, Lobachevsky State University of Nizhni Novgorod, 603022 Nizhny Novgorod, Russia}
\affiliation{Skobeltsyn Institute of Nuclear Physics, Lomonosov Moscow State University, 119991 Moscow, Russia, Russia}
\affiliation{Moscow Technical University of Communication and Informatics (MTUCI), 111024 Moscow, Russia}

\author{N. V. Klenov}
\affiliation{National University of Science and Technology MISIS, 119049 Moscow, Russia}
\affiliation{Faculty of Physics, Lomonosov Moscow State University, 119991 Moscow, Russia}

\begin{abstract}
In this article, we consider designs of simple analog artificial neural networks based on adiabatic Josephson cells with a sigmoid activation function. A new approach based on the gradient descent method is developed to adjust the circuit parameters, allowing efficient signal transmission between the network layers. The proposed solution is demonstrated on the example of the system implementing XOR and OR logical operations.
\end{abstract}


\date{\today}
\maketitle

\section{Introduction}

A distinctive feature of the current era of information technology evolution is the widespread development and implementation of artificial intelligence (AI) \cite{zbontar2016stereo, tolosana2018exploring, kaya2019deep, ruiz2020off, wang2021deep, ilina2022survey}. In order to effectively solve a number of tasks, specialised hardware implementation of AI systems is required \cite{le202364, modha2023neural}. The most popular and exciting at the moment are the so-called neuromorphic chips or neuromorphic processors. In this field, world giants such as Intel (Loihi~1 and Loihi 2) and IBM (TrueNorth, NorthPole) have made their mark. In addition to neuromorphic processors, there are machine learning processors (Intel Movidius Myriad 2, Mobileye EyeQ) designed to accelerate data processing (video, machine vision, etc.) and tensor processors (Google TPU, Huawei Ascend, Intel Nervana NNP) designed to accelerate arithmetic operations. While the latter two types have been successfully implemented in modern hardware platforms (smartphones, cloud computing, etc.), neuromorphic processors, despite their potential, are unfortunately not yet widespread and remain mostly at the laboratory production and testing stage \cite{kumar2013introducing,prezioso2015training, bose2017stable, davies2018loihi, cheng2018spiking, jeong2018memristor, debole2019truenorth,arute2019quantum, berggren2020roadmap, wan2022compute}.

\begin{figure}[b!]
\center{\includegraphics[scale=0.5]{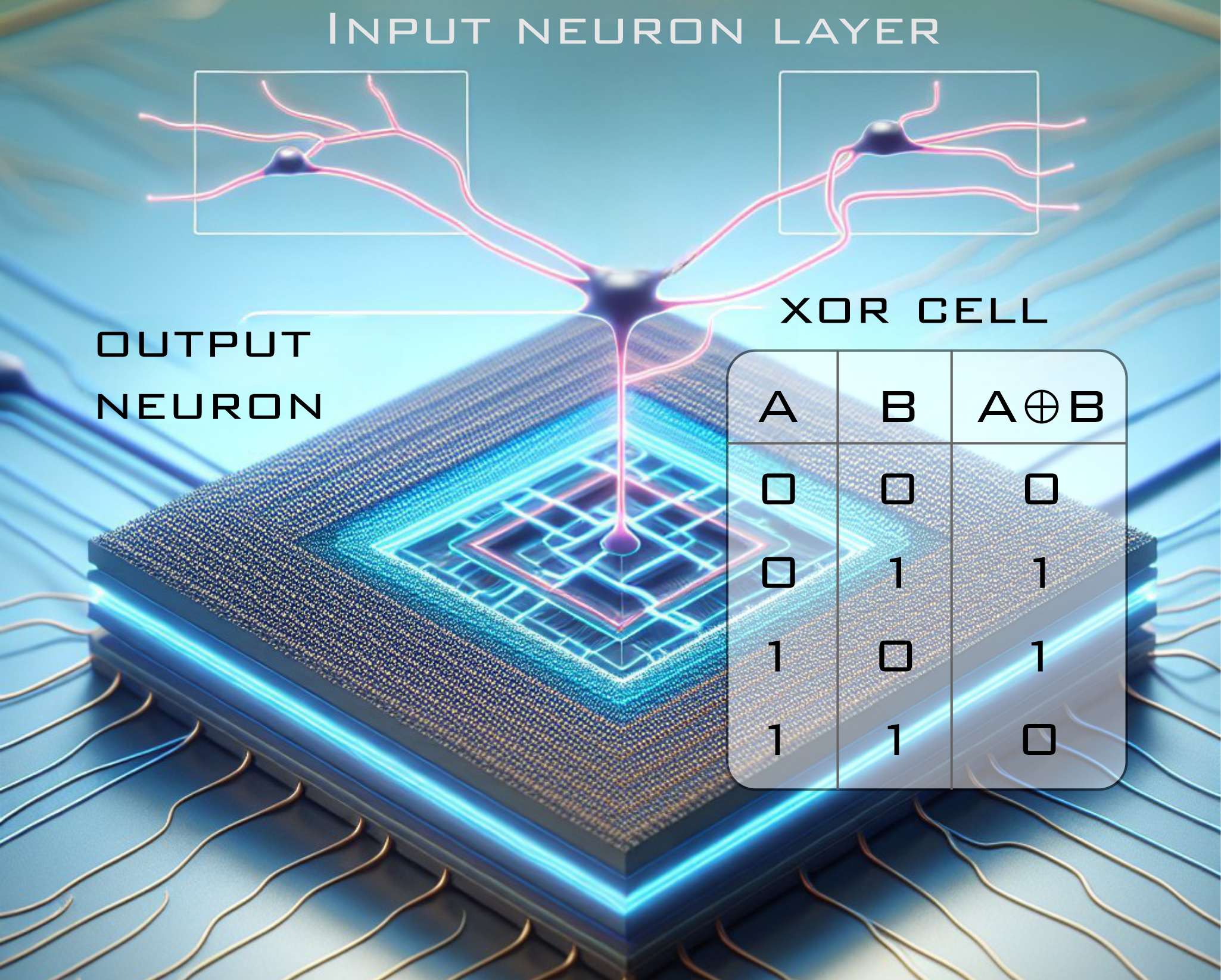}}
\caption{
OpenAI's DALLE 3 prompt-generated image of a superconducting neural network simulating an XOR operation.
}
\label{fig0}
\end{figure}

There are a number of post-Moore technology platforms that enable the realisation of AI technologies at the hardware level, promising advances in performance and/or energy efficiency. Optical neuromorphic networks are an excellent example \cite{feldmann2019all, jha2021photonic} of energy-efficient systems with high performance. 
Photonic-superconducting interfaces \cite{singh2014superconductor, fan2018superconducting, gu2017microwave, berman2006superconducting} and other hybrid optical-superconducting neural networks \cite{shainline2017superconducting, shainline2018circuit, shainline2019superconducting, schneider2022supermind} were once a major milestone in the development of this field of applied science. These systems used light pulses to transmit signals and superconducting circuits based on quantum interferometers to process and store information. Superconducting elements are known for their high energy efficiency \cite{crotty2010josephson, russek2016stochastic, schneider2018ultralow, toomey2020superconducting, ishida2021superconductor, zhang2021brain, semenov2021new, casaburi2022superconducting, feldhoff2024short}. 
In the context of modern data centres that require massive cooling, superconductor-based hybrid computers  may become quite competitive players. It is also worth noting that quantum computers \cite{siddiqi2021engineering, vozhakov2022state, Calzona_2023} are now being developed on the basis of superconductor technology. Therefore, the creation of superconducting neuromorphic chips capable of hybridisation with quantum computers (QCs) seems very reasonable. An example might be qubit spectrum detection of a QC's output signal, or a QC's calculation of synaptic weights for an externally tunable artificial neural network. This study focuses on the optimisation of superconducting basic elements and their interconnections specified for superconducting logic gates in neuromorphic systems (Figure~\ref{fig0}).

It is also necessary to mention here imitations of neural activity of living tissues with the help of superconducting electronics using Josephson contacts \cite{crotty2010josephson, russek2016stochastic, segall2017synchronization, schneider2018ultralow, toomey2020superconducting, zhang2021brain, feldhoff2021niobium, goteti2021superconducting, chalkiadakis2022dynamical, schneider2022supermind, schegolev2023bio, crotty2023biologically, feldhoff2024short}. These works demonstrate the operation of bio-inspired neurons (capable of reproducing basic biological patterns of nervous activity, such as excitability, spiking and bursting) and synapses, as well as simple neural networks. The possibility of using Josephson circuits for modeling and simulating the work of neurons and tissues, as well as for more applied tasks (e.g., recognition), will allow reaching a new level of performance (computational and modeling speed, energy efficiency) of spiking neural networks.

Previously, we presented the concept of an adiabatic interferometer-based superconducting neuron \cite{schegolev2020learning,  bastrakova2021dynamic, ionin2023experimentalS}, capable of operating in classical and quantum modes with ultra-low energy dissipation per operation (in the zJ range) \cite{takeuchi2020directly, khazali2020fast, ayala2020mana, yamazaki2021compact, setiawan2021analytic, bastrakova2022superconducting, pashin2023bifunctional, mizushima2023adiabatic}. The development of an adiabatic perceptron requires the realisation of a large number of connections between neurons via superconducting synapses \cite{schegolev2020learning}. Good synapses for perceptron-type networks should have the following important properties: a wide range of weights (both negative and positive, as well as zero), low noise, signal type preservation (high linearity), and circuit simplicity (as few components as possible). Based on these requirements we used the synapse scheme first presented in \cite{bakurskiy2020controlling}.

Combining these elements into an analog network implies the generally difficult task of studying the complex nonlinear dynamics of the system. We propose a solution to this problem and demonstrate the results on the example of a three-neuron network simulating an XOR and OR logic gates.

\section{The model for two coupled adiabatic neurons}

\begin{figure}[h!]
\center{\includegraphics[scale=0.45]{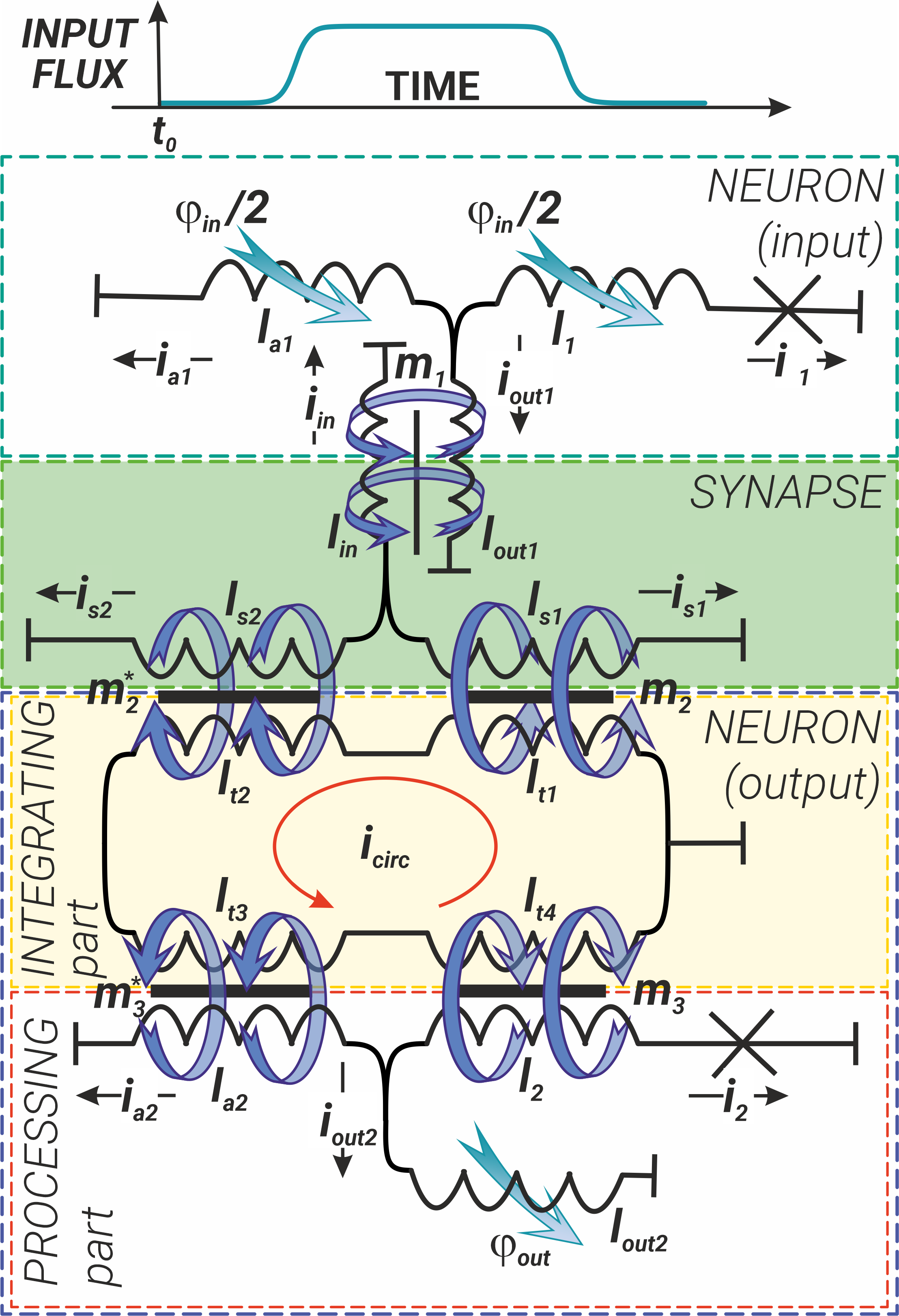}}
\caption{
Schematic representation of two coupled $S_c$-neurons (input -- in cyan box and output -- in navy blue box), connected through the inductive synapse (in the green box) and the coupler -- integrating part of the output neuron (in the light yellow box). Processing part of the} output neuron is highlighted by red box. Black or red arrows and blue curled arrows indicate currents and corresponding magnetic fluxes, respectively.

\label{fig1}
\end{figure}

Before the simulation of the superconducting logic element, we have considered the system of two coupled $S_c$-neurons having sigmoid activation function. These basic elements are the superconducting interferometers connected by the inductive synapse, see Figure~\ref{fig1}. The formation of the activation functions (flux-to-flux transformations) on individual $S_c$-neurons has been previously studied in detail both in classical \cite{ schegolev2020learning,  bastrakova2021dynamic, ionin2023experimentalS} and quantum modes \cite{bastrakova2022superconducting,pashin2023bifunctional}. Here we consider the interaction between different parts of the system. We choose an inductive synapse instead of the Josephson one \cite{njitacke2022extremely} because of its absolute linearity of the transfer characteristic and a wide dynamic range \cite{bakurskiy2020controlling, schegolev2022tunable}.

The $S_c$ neurons (areas outlined by the cyan and navy blue dashed lines in Figure~\ref{fig1}) are designed according to the integrating-and-processing principle: the integrating part (IP) collects or $integrates$ all input signals, while the processing part (PP) processes this input signal and generates an output signal. Generally $S_c$-neuron consists of three branches (its processing part): two of them (the branches with the inductance $l_{out1,2}$ and the branch with the Josephson junction and the inductance $l_{1,2}$) form the circuit of the so-called quantron. The third branch with a single inductance $l_{a1,2}$ shunts the quantron circuit. In the Figure~\ref{fig1} the IP of the output neuron is highlighted by light yellow box and is formed by a so-called coupler -- an inductive ring ($l_{t1,2,3,4}$) that collect output flux from input neuron(s) (the IP of the input neuron is not shown in the Figure~\ref{fig1}). The signal in the form of magnetic flux flows from the neuron's IP to the neuron's PP through the inductances $l_{1,2}$ and $l_{a1,2}$. The inductance $l_{out1,2}$ is used to transmit the magnetic flux from the $S_c$-neuron to subsequent element (in our case the input neuron transmits its signal to the inductive synapse).

The inductive synapse (green box in Figure~\ref{fig1}) in turn also has three branches: the input branch (containing the inductance $l_{in}$) is responsible for signal reception, and the branches containing the tunable kinetic inductances $l_{s1}$ and $l_{s2}$ \cite{annunziata2010tunable, splitthoff2022gate, klenov2019periodic, bakurskiy2020controlling} provide its further transmission. Synapse adjustment is realised by external magnetic or spin-current influence (not shown in Figure~\ref{fig1}). By changing the values of the inductances $l_{s1}$ and $l_{s2}$ one can vary the weight of the synapse. 

In the following, all inductances are normalised to the characteristic Josephson inductance of the output neuron Josephson junction, $\Phi_0/2\pi I_{C_{2}}$, where $I_{C_{2}}$ is the critical current of this junction. Magnetic fluxes are normalised to the magnetic flux quantum, $\varphi = 2 \pi \Phi / \Phi_{0}$, $\Phi_0 = h/2e$.

The input signal ($\varphi_{in}$) has been set in the form of a smoothed trapezoid, which makes it possible to take into account both the rising (rise time) and falling (fall time) phases of the signal. The duration of the plateau section can also be controlled:
\begin{equation}
\begin{aligned}
    \varphi_{in}(t) = A_{in}\cdot \left\{\frac{1}{1 + \exp(-2D(t - t_1))} + \right. \\ \left.
    \frac{1}{1 +  \exp(2D(t - t_2))}\right \} - A_{in}.
    \label{eq:phi_in}
\end{aligned}
\end{equation}
The parameters $A_{in}$ and $D$ set the level and the rise/fall rate of the input magnetic flux respectively. As it is shown in \cite{bastrakova2021dynamic}, the input signal in the form of (\ref{eq:phi_in}) allows to obtain the sigmoid transfer function of the $S_c$-neuron for certain values of the inductances.

The circuit shown in Figure~\ref{fig1} is described by the following system of equations:
\begin{equation}
 \begin{cases}
   i_{a1}+i_{1}+i_{out1} = 0,
   \\
   \varphi_{1}-\frac{\varphi_{in}}{2} + i_{1} l_{1} = i_{out1} l_{out1} + m_{1} i_{in},
   \\
   \varphi_{1}-\frac{\varphi_{in}}{2} + i_{1} l_{1} = i_{a1} l_{a_{1}} + \frac{\varphi_{in}}{2},
   \\
   i_{in} + i_{s1} +i_{s2} = 0,
   \\
   i_{s1} (l_{s1}+l_{p}) + m_{2} i_{circ} = i_{s2} (l_{s2}+l_{p}) - m^*_{2} i_{circ},
   \\
   i_{s1} (l_{s1}+l_{p}) + m_{2} i_{circ} = -i_{in} (l_{in} + l_{p}) + m_{1} i_{out1},
   \\
   i_{circ} \sum_{j=1}^{4} l_{tj} = i_{2} m_{3} - i_{a2} m^*_{3} -i_{s1} m_{2} + i_{s2} m^*_{2},
   \\
   i_{2} + i_{a2} + i_{out2} = 0,
   \\
   \varphi_{2} - m_{3} i_{circ} + i_{2} l_{2} = i_{out2} l_{out2},
   \\
   \varphi_{2} - m_{3} i_{circ} + i_{2} l_{2} = i_{a2} l_{a_{2}} + m^*_{3} i_{circ}.
 \end{cases}
 \label{eq:System}
\end{equation}
Here $\varphi_{1,2}$ are the superconducting phase drops at the Josephson junctions of the input and output neurons, $l_p$ is an additional non-adjustable (parasitic) inductance in this circuit, which is not explicitly shown in the Figure~\ref{fig1}, but is taken into account in our calculations. The currents $i_{a1}$, $i_{1}$ and $i_{out1}$ are the currents flowing through the corresponding inductances in the input neuron $l_{a1}$, $l_{1}$ and $l_{out1}$. The currents $i_{in}$, $i_{s1}$ and $i_{s2}$ are the currents in the synapse, flowing through $l_{in}$, $l_{s1}$ and $l_{s2}$. The currents $i_{s1}$ and $i_{s2}$ induce the circulating current $i_{circ}$ in the integrating part of the output neuron. The circulating current in turn induces currents in the processing part of the output neuron $i_2$, $i_{s2}$ and $i_{out2}$ which flow through the inductances $l_{a2}$, $l_{2}$ and $l_{out2}$. All currents in (\ref{eq:System}) are normalised by $I_{C_{2}}$. The parameters $m_{k}$ and $m_{k}^{*}$ are mutual inductance coefficients in transformer elements ($k=1,2,3$), which are considered equal to the average values of the inductances constituting the corresponding transformers.

It can be shown that the currents in the proposed circuit (Figure~\ref{fig1}) have a simple relationship with the phases of the Josephson junctions and the external flux:
\begin{equation}
   i_{\gamma} = - \kappa_{\gamma}^{(1)}\varphi_{1} - \kappa_{\gamma}^{(2)}\varphi_{2} - \kappa_{\gamma}^{(in)}\varphi_{in}.
    \label{eq:2}
\end{equation}
Here all coefficients $\kappa_{\gamma}^{(1)}$, $\kappa_{\gamma}^{(2)}$ and $\kappa_{\gamma}^{(in)}$ are obtained from the system of equations (\ref{eq:System}) and represented in terms of inductances according to Figure~\ref{fig1}. The subscript $\gamma = 1, 2, in, s, a, out$ of the coefficients indicates the current ($i_{1}$, $i_{2}$, $i_{in}$, and $\Delta i_s = i_{s1} - i_{s2}$, respectively) to which they belong. The superscript in the formula~(\ref{eq:2}) takes the values of the corresponding phases of the junctions or the input magnetic flux. The analytical expressions for these coefficients are bulky, so they are given in the Supplementary Material. 

Note that due to $\varphi_{1}$, $\varphi_{2}$ complex dependence on $\varphi_{in}$, in fact, the currents $i_{\gamma}$ are not linear in any of them. The non-linearity of the system comes from the Josephson junctions, whose currents, $I_n$ (where the subscript $n = 1, 2$ is the index of the junction), can be written as
\begin{equation}
  I_{n} = \frac{\hbar}{2e}C_{n}\ddot{\varphi_{n}} + \frac{\hbar}{2eR_{n} } \dot{\varphi_{n}} + I_{C_{n}}\sin\varphi_{n}.
    \label{EQ:dynamic}
\end{equation}
in the frame of the well-known resistively shunted junction model with capacitance (RSJC) \cite{Stewart1968}. 

Here we consider an energy efficient circuit consisting of tunnel superconducting - insulator - superconducting (SIS) Josephson junctions with a high normal state resistance $R_n$, so that the second term in the equation~(\ref{EQ:dynamic}) becomes negligibly small and, as modelling shows, does not contribute significantly to the overall dynamics of the system, therefore can be safely omitted.

After normalisation (\ref{EQ:dynamic}) by $I_{C_{2}}$, the equations take the following form:
\begin{equation}
   c_{n}\ddot{\varphi_{n}} + i_{C_{n}}\sin\varphi_{n} = i_{n},
    \label{EQ:dynamic_nont}
\end{equation}
where $i_{C_n} = I_{C_n} / I_{C_{2}}$ is a dimensionless critical current; $t_{C} = \sqrt{\frac{\hbar C_2}{2 e I_{C_2}}}$ is a characteristic time and $\tau = t/t_C$ is a dimensionless time; $c_{n} =  C_n / C_2$ is a dimensionless capacity. Note that such systems of interacting neurons can also be considered within the framework of the Hamiltonian formalism. As an example, in the Supplementary Materials, we present the derivation of the Hamiltonian of the system shown in Figure~\ref{fig1}. This approach is quite simple and convenient in the case of scaling the circuit to a larger number of layers in a neural network, as well as for numerical modeling of nonlinear dynamics and further study of the quantum mode of operation of the circuit \cite{bastrakova2022superconducting, pashin2023bifunctional}, including taking into account the influence of environments.

Solution of the system of equations (\ref{EQ:dynamic_nont}) gives the transfer characteristics of the input and output neurons as a response to the input magnetic flux (\ref{eq:phi_in}). Previous studies of single $S_c$-neurons \cite{bastrakova2021dynamic} have shown that the sigmoid activation function can be realised under the following condition: $l_n< \sqrt{l_{out n}^2+1}-l_{out n} \equiv l_{n}^{*}$ and $l_{a n}=l_n+1$.
Hence, as in the single-neuron case, we will consider values of inductances $l_n<l_{n}^{*}$ at which there are no plasmonic oscillations in the output characteristics of the first (input) and the second (output) neurons.

\begin{figure}[h!]
\center{\includegraphics[scale=0.3]{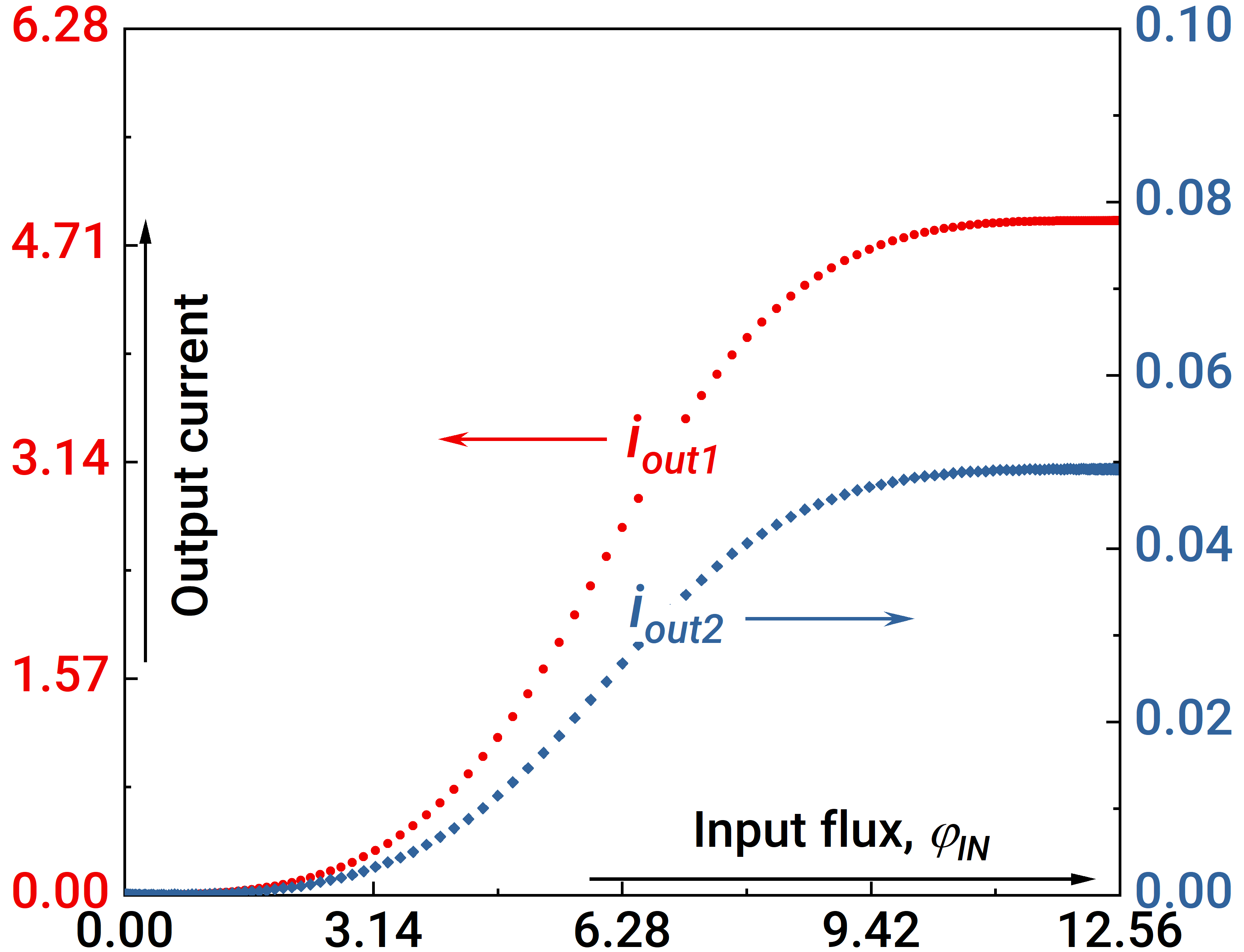}}
\caption{
Activation functions for the first (input) and second (output) neurons. 
The "red" curve corresponds to $i_{out1}$, the "blue" curve corresponds to $i_{out2}$. Parameters of the system: $l_{1,2}=0.2$, $l_{out1}=l_{out2}=1$, $l_{a1,a2}=l_{1,2}+1$, $l_{in}=1$, $l_{t1}=l_{t2}=0.1$, $l_{t3}=l_{t4}=1$, $l_{s1}+l_{s2}=1$, $l_{s1}-l_{s2}=0.9$.
}
\label{fig2}
\end{figure}

\begin{figure}[h!]
\center{\includegraphics[scale=0.45]{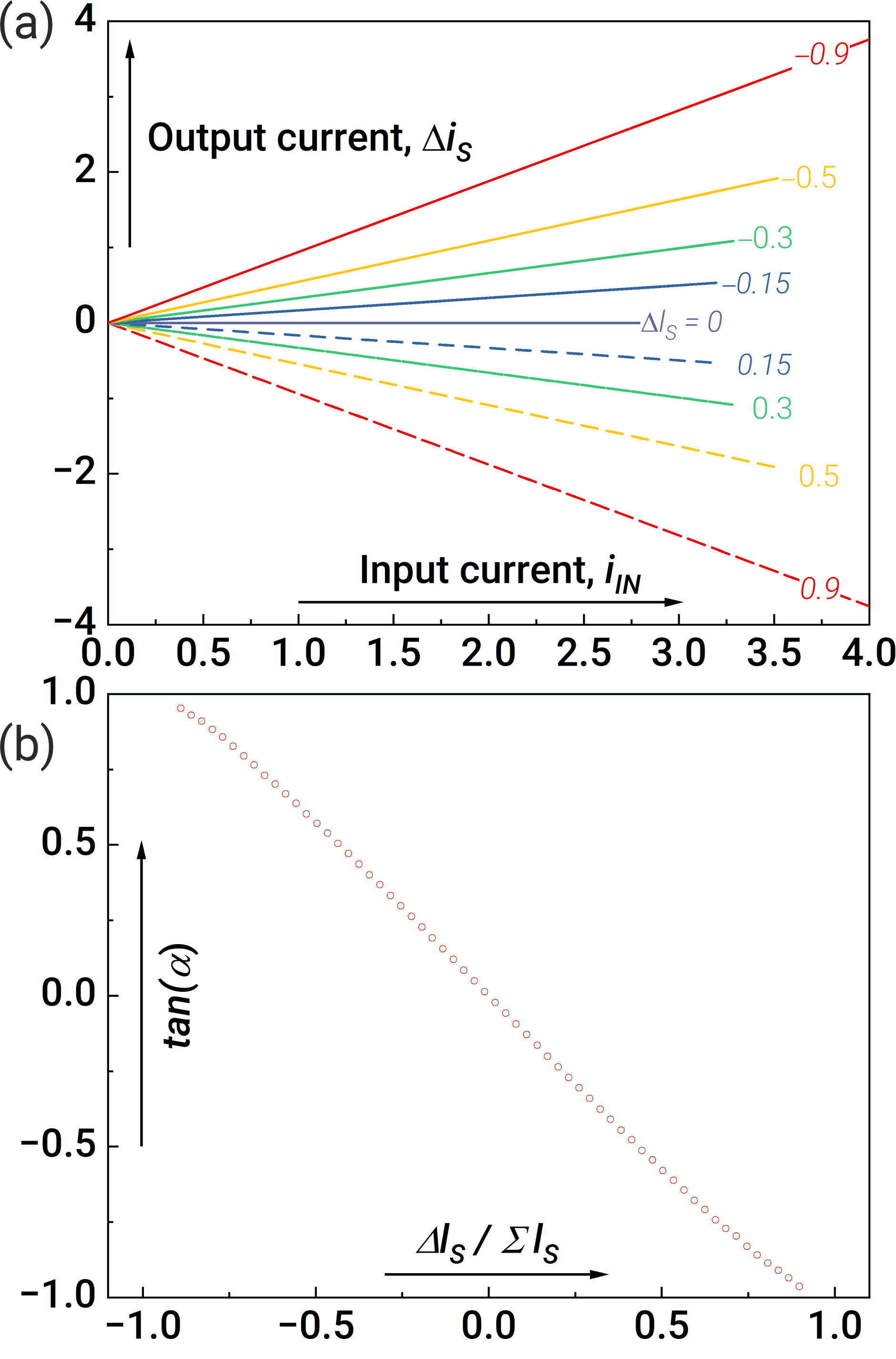}}
\caption{Synaptic weights without system parameters optimisation: (a) dependence of the output current from the synapse ($\Delta i_s = i_{s1} - i_{s2}$) as a function of the input current ($i_{in}$) and (b) calculations for the dependence of the slope angle $\alpha$ on $\Delta l_s = l_{s1} - l_{s2}$. Parameters of the system: $l_{1,2}=0.2$, $l_{out1}=l_{out2}=1$, $l_{a1,a2}=l_{1,2}+1$, $l_{in}=1$, $l_{t1}=l_{t2}=0.1$, $l_{t3}=l_{t4}=1$, $l_{s1}+l_{s2}=1$.
}
\label{fig3}
\end{figure}

In the first step of the analysis we assume that the coupler inductances should be equal: $l_{t1}=l_{t2}$ and $l_{t3}=l_{t4}$. Figure~\ref{fig2} illustrates the formation of sigmoid activation functions for the input and output neurons under this assumption. It is seen that, the current at the output neuron (blue curve in Figure~\ref{fig2}) drops by two orders of magnitude; this drawback reflects the difficulty of practical system implementation. It is also necessary to obtain the synapse weights that are at least in the range from -1 to +1, which turns out to be impossible in some situations (see Figure~\ref{fig3}).

The above issues imply the need for parameter optimisation. In the next part of the paper, we propose this procedure, which can be generalised to the case of large computing systems.

\section{Formulation and solution of the optimisation problem}

We consider the optimisation problem of a system of two coupled neurons from the point of view of solving the two problems of synapse weights and neuron response magnitudes mentioned above. However, a closer look at these problems reveals that they are closely related: achieving higher values of weights can potentially increase the response magnitude of the output. Therefore, further actions will be aimed at finding a functional describing the synapse weight as a function of the system parameters and finding its extrema using the gradient descent method. As such a functional, we consider the slope angle of the synapse characteristic $\alpha$, which can be expressed analytically in the following form:
\begin{equation}
\begin{aligned}
\tan{ \alpha} = \frac{d \Delta i_{s}}{dt} \Big/ \frac{d i_{in}}{dt} = \\
\frac{\kappa^{(1)}_{ s} \dot \varphi_1+\kappa^{(2)}_{s} \dot \varphi_2+\kappa^{(in)}_{s} \dot \varphi_{in}}{\kappa^{(1)}_{in} \dot \varphi_1+\kappa^{(2)}_{in} \dot \varphi_2+\kappa^{(in)}_{in} \dot \varphi_{in}},
\label{alpha1}
\end{aligned}
\end{equation}
When using the gradient descent method, it is necessary to solve a system of differential equations (\ref{EQ:dynamic_nont}) at each step, which is the main computational complexity due to the large number of varied system parameters. To overcome this difficulty, we propose several simplifications.

Since dynamic processes in the system are associated with changes in the input flux and, moreover, take place exactly at the rise/fall time intervals, and the dependence $\Delta i_s(i_{in})$ is linear, it is sufficient to determine the value of the angle (\ref{alpha1}) at the inflection point $t_1$ when $\ddot\varphi_{in} (t_1)=0$. Besides, $\ddot\varphi_{1} (t_1)=\ddot\varphi_{2} (t_1)=0$, due to sigmoid activation function. By using this approximation, we obtain the system of equations for $\dot\varphi_{1} (t_1)$ and $\dot\varphi_{2} (t_1)$:
\begin{equation}
\label{sys2}
 \begin{cases}
\dot\varphi_{1} (t_1)=\dot\varphi_{in}\eta^{-1}(\kappa\kappa_{2}^{(in)}-\kappa_{1}^{(in)}\kappa_{2}^{(2)}-\kappa_{1}^{(in)}i_{C_2}\cos{(\varphi_2(t_1))}),\\
\dot\varphi_{2} (t_1)=\dot\varphi_{in} \eta^{-1}(\kappa\kappa_{1}^{(in)}-\kappa_{2}^{(in)}\kappa_{1}^{(1)}-\kappa_{2}^{(in)}i_{C_1}\cos{(\varphi_1(t_1))}),
 \end{cases}
\end{equation}
where $\eta \equiv -\kappa^{2}+(i_{C_1}\cos{(\varphi_1(t_1)})+\kappa_{1}^{(1)}) \cdot (i_{C_2}\cos{(\varphi_2(t_1))} +\kappa_{2}^{(2)})$, remind that $\kappa \equiv \kappa_{1}^{(2)} = \kappa_{2}^{(1)}$, and the values of $\varphi_{1} (t_1)$ and $\varphi_{2} (t_1)$ can be found from:
\begin{equation}
\label{sys3}
 \begin{cases}
i_{C_1}\sin{(\varphi_1(t_1))} = -(\kappa \varphi_{2} (t_1) + \kappa_{1}^{(1)} \varphi_{1} (t_1) + \kappa_{1}^{(in)} \varphi_{in} (t_1)),  \\
i_{C_2}\sin{(\varphi_2(t_1))} = -(\kappa \varphi_{1} (t_1) + \kappa_{2}^{(2)} \varphi_{2} (t_1) + \kappa_{2}^{(in)} \varphi_{in} (t_1)).
 \end{cases}
\end{equation}

By substituting the obtained values of $\dot\varphi_{1} (t_1)$ and $\dot\varphi_{2} (t_1)$ into the expression (\ref{alpha1}), we obtain an explicit form for $\alpha$ depended on all system parameters. This allows us to implement the gradient descent method to maximise the angle $\alpha$ without directly calculating the dynamics (\ref{EQ:dynamic_nont}). Similar approach allows us to quickly optimise the parameters to maximise the current at the output neuron by using (\ref{eq:2}).

\begin{figure}[h!]
\center{\includegraphics[scale=0.5]{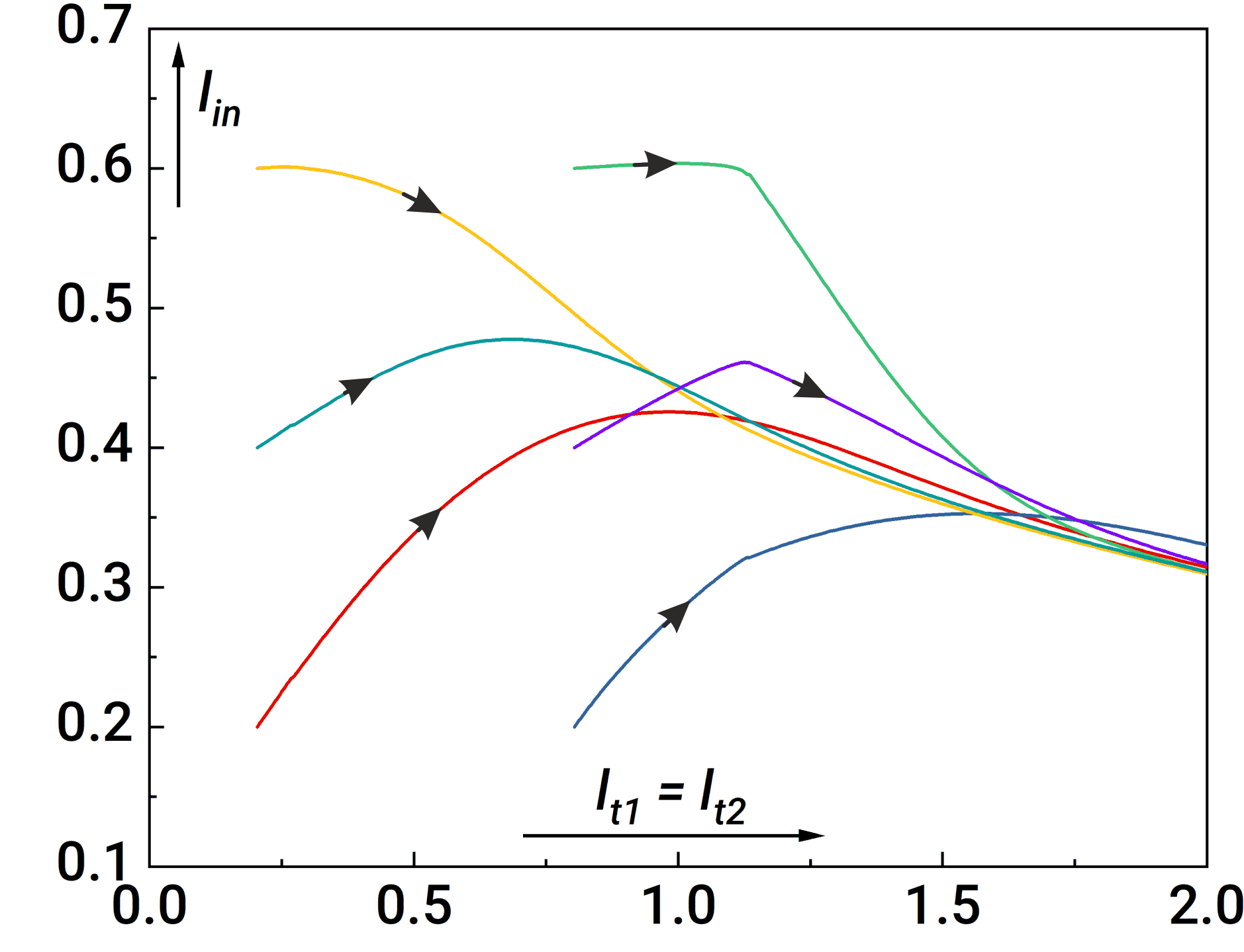}}
\caption{Gradient descent trajectories for $\max\alpha$ maximisation for different initial parameters (shown by different colors), projected onto the plane $\{l_{in};l_{t1,t2}\}$.
} 
\label{fig4}
\end{figure}

A visualisation of this method for different initial parameters is shown in Figure~\ref{fig4}. We selected several initial sets of system inductances, for which $\alpha$ was calculated using (\ref{alpha1}) and maximised based on the gradient descent method. The angle $\max\alpha$ is non-monotonic with respect to the system parameters and has several local maxima. In the  Figure~\ref{fig4} we show a section for several trajectories along which the angle $\max\alpha$ is maximised in the subspace of inductances $l_{in}$ and $l_{t1} = l_{t2}$, where the arrow indicates the path from their initial values to the optimal ones. It is seen that all curves converge at $l_{t1, 2} \rightarrow 2$ (which was chosen as an upper boundary value for the inductances $l_{t1...4}$) and $l_{in} \rightarrow 0.3$, where a certain local maximum of optimisation is reached for $\max\alpha$, and therefore, for the achievable synapse weights in our system.

\begin{figure}[t]
\center{\includegraphics[scale=0.5]{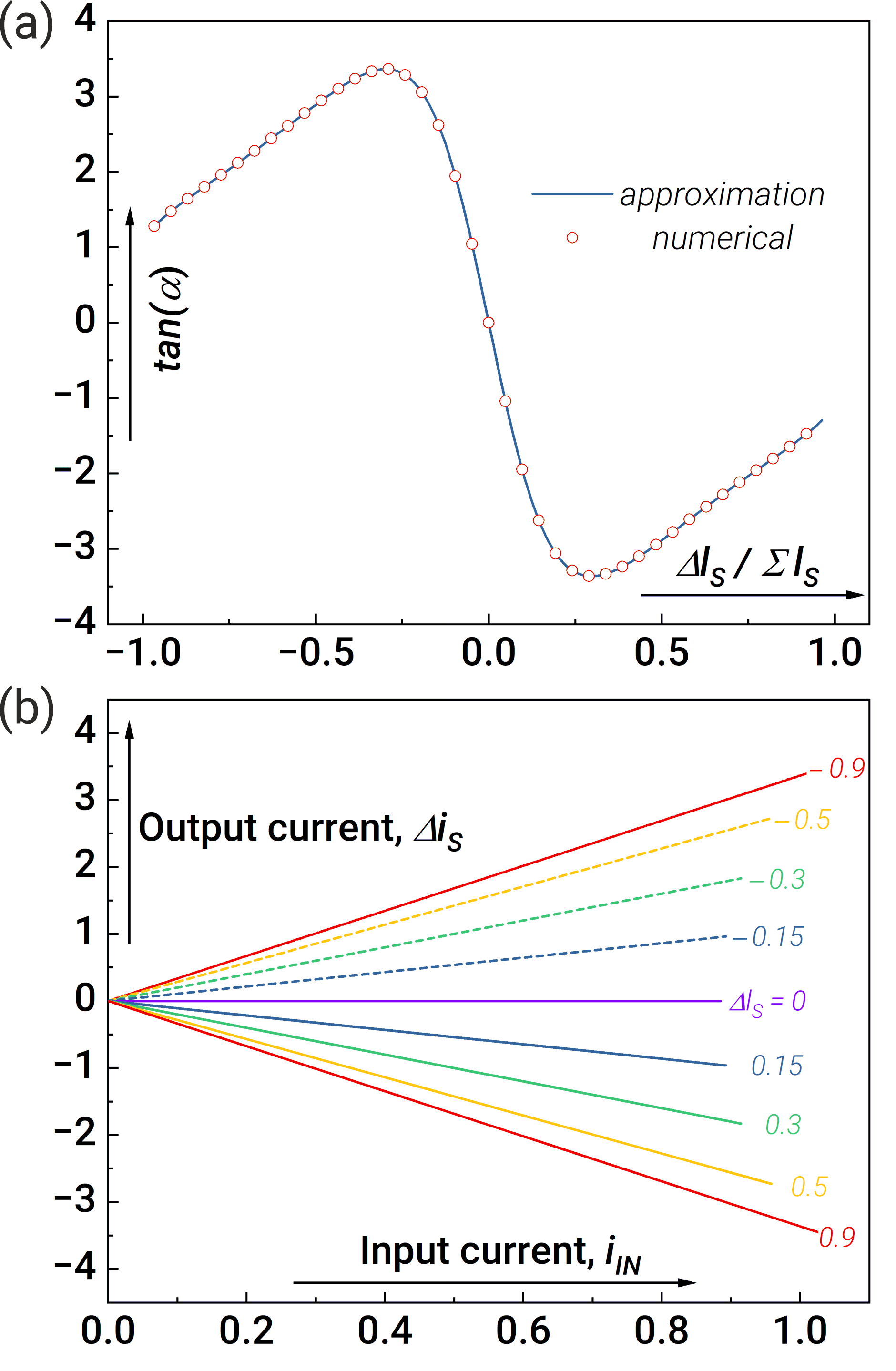}}
\caption{
Demonstration of synapse capabilities after system parameters optimisation: (a) comparison of numerical and analytical calculations for the dependence of the slope angle $\alpha$ on $\Delta l_s$ and (b) dependence of the output current from the synapse $\Delta i_s$ as a function of the input current $i_{in}$.
The blue line shows the result of the approximate calculation using Eqs.~(\ref{alpha1}) -- (\ref{sys3}). The red circles show the result of the exact numerical calculation of the dynamics from Eq.~(\ref{EQ:dynamic_nont}).
Parameters of the system are: $l_{1,2}=0.1$, $l_{in}=0.3$, $l_{t1}=l_{t2}=2$, $l_{t3}=l_{t4}=0.1$, $l_{out1,2}=0.1$, $l_{s1}+l_{s2}=3$.
}
\label{fig5}
\end{figure}

Figure~\ref{fig5}a shows $\text{tan}(\alpha)$ dependence on the inductance difference $\Delta l_s$ for optimal system parameters found by the gradient descent method. The good agreement between the results obtained from the exact calculation of Eq.~(\ref{EQ:dynamic_nont}) (the red circles) and by using Eqs.~(\ref{alpha1}) -- (\ref{sys3}) (the blue line) indicates the validity of the approximations used. Dependencies of the synapse output current $\Delta i_s$ on the input current $i_{in}$ for different values of $\Delta l_s$ are shown in the Figure~\ref{fig5}b.

The proposed method allows to abandon the solution of the Hamiltonian system, which is a time-consuming computational task. We reduce the optimisation problem to solving a set of algebraic equations, which significantly reduces the computational time. This approach is promising from the point of view of scaling neural networks and calculating their optimal configuration parameters.

The obtained results demonstrate that the gradient descent method can be used to optimise the parameters of a synapse connecting two neurons, 
Extending the applicability of the method to more complex systems consisting of a larger number of neurons and synapses is also possible, but may require additional assumptions related to mutual influence of neurons on each other (localisation approximation). Hence, the challenge of optimisation of the parameters of a large neural network will be reduced to solving local problems of finding functionals similar to Eq.~(\ref{alpha1}) and then fine tuning the found solutions by the gradient descent method in a multi-parameter space.

\section{Circuit structure optimisation}

\begin{figure}[h!]
\center{\includegraphics[scale=0.45]{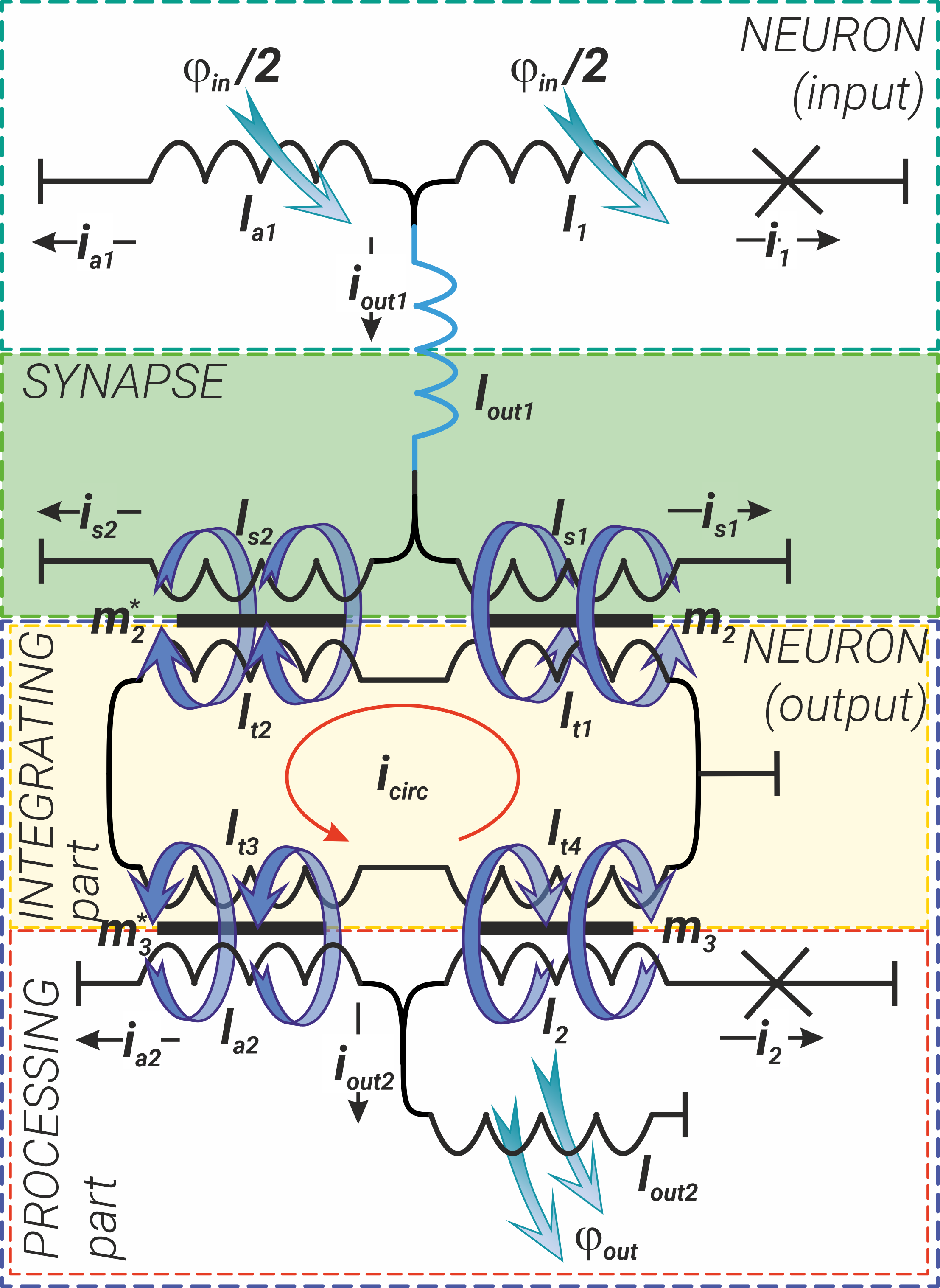}}
\caption{
Schematic representation of the modified coupling between two $S_c$-neurons: the transformer consisting of inductances $l_{out1}$ and $l_{in}$ coupling the input neuron and the synapse (see  Figure \ref{fig1}) is replaced by a direct coupling via the inductance $l_{out1}$ only.
}
\label{fig6}
\end{figure}

The performed parameter optimisation does not eliminate the signal level drop at the output neuron in the considered circuit design, Figure~\ref{fig1}. To overcome this problem, we are developing a modification of the circuit in which the magnetic connection between the input neuron and the synapse is replaced by a galvanic connection, see Figure~\ref{fig6}.

\begin{figure}[h!]
\center{\includegraphics[scale=0.52]{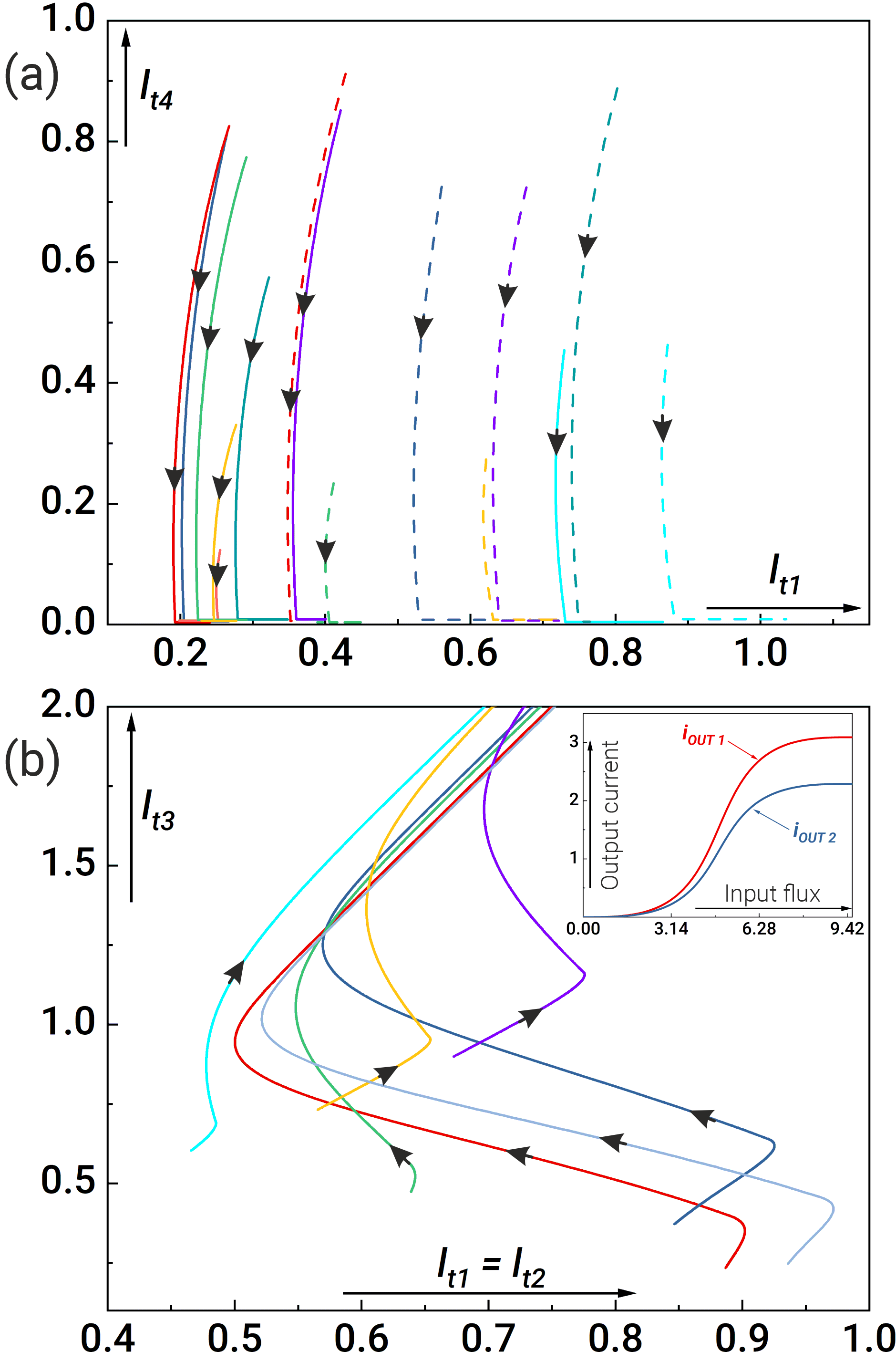}}
\caption{Projection of gradient descent trajectories for different initial parameters (each curve corresponds to its own initial values) (a) on axes $l_{t4}$ and $l_{t1}$  and (b) on axes $l_{t3}$ and $l_{t1}=l_{t2}$. The inset of figure (b) shows the transfer characteristics of the input and output neurons ($i_{out1,2}(\varphi_{in})$) with the following system parameters: $l_{1,2}=0.1$, $l_{t1}=l_{t2}=0.7$, $l_{t3}=1.5$, $l_{t4}=0.1$, $l_{out1,2}=0.1$, $\Sigma l_{s}=3$, $\Delta l_s =1.53$. 
}
\label{fig7}
\end{figure}

Within the framework of the proposed approach, gradient descent was applied to the modified scheme to solve the optimisation problems. The analysis of the system showed that the main parameters responsible for the current at the output neuron are the coupling inductances $l_{tj}$ (where $j=1,2,3,4$). Figure~\ref{fig7}a shows that we need to minimise the value of the inductance $l_{t4}$ connecting the coupler to the Josephson arm in the output neuron. The direction of the arrows shows the path of the trajectory (from the initial value to the optimal one) for maximising the angle $\max\alpha$ during the gradient descent execution. Figure~\ref{fig7}b shows the calculation for optimisation of the remaining coupler inductances. It can be seen that all trajectories tend to the values $l_{t3}\rightarrow 2$ and $l_{t1,2}\rightarrow 0.7$. We calculate the activation functions of the neurons shown in the inset of Figure~\ref{fig7}b using these values. By application of the optimisation approach, we are able to significantly increase the current at the output neuron, which is important for the practical implementation of such systems.

\begin{figure}[t]
\center{\includegraphics[scale=0.5]{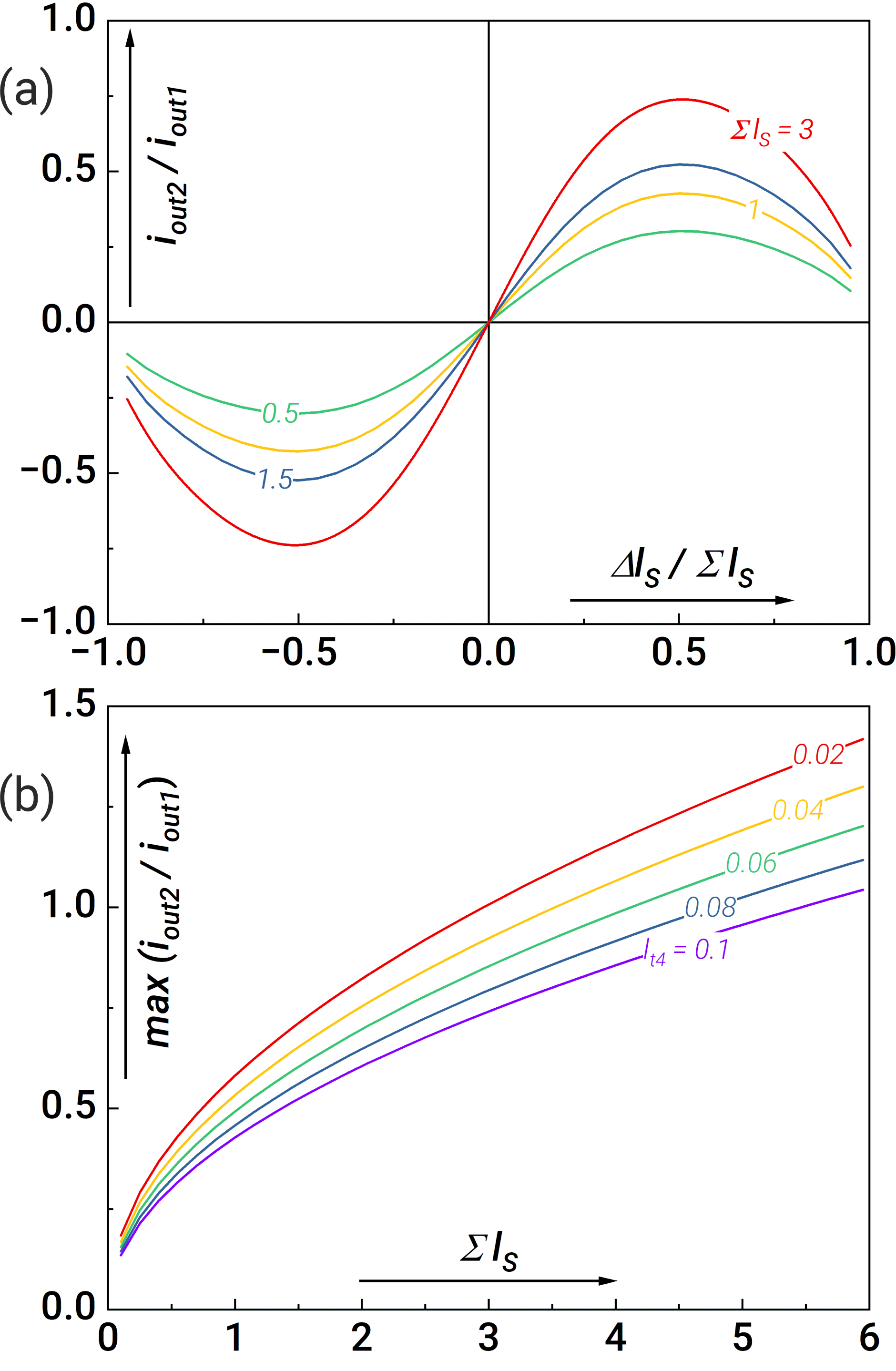}}
\caption{
(a) Dependence of the current ratio $i_{out2}/i_{out1}$ at the moment when the input flux reaches a plateau at $t= (t_1+t_2)/2$ on the normalised difference of the inductances of the synaptic arms for $l_{t4}=0.1$ and different values of the inductance sum.
(b) Maximum ratio between output currents $i_{out2}$ and $i_{out1}$ for different values of $l_{t4}$ in dependence on the inductance sum, $\Sigma l_s$. 
Other system parameters are follows: $l_{1,2}=0.1$, $l_{t1}=l_{t2}=0.7$, $l_{t3}=1.5$,  $l_{out1,2}=0.1$.
}
\label{fig8}
\end{figure}

After re-optimisation of the parameters, we re-examine the synaptic weights. We analyze the dependence of the current ratio $i_{out2}/i_{out1}$ on $\Delta l_S$ where the input flux reaches a plateau at $t = (t_1 + t_2) / 2$, see Figure~\ref{fig8}a. It can be seen, that we can adjust the sum of the inductances such that the values of the out currents at the input and output neurons coincide, see Figure~\ref{fig8}b. Note that the output current of the output neuron can even exceed the output current of the first neuron at small values of $l_{t4}$. Thus, depending on the technological limitations, it is possible to obtain the maximum response at the output layer of the neurons.

\section{Analog implementation of the XOR and OR logic elements}

\begin{figure}[h!]
\centering
\includegraphics[width=0.95\linewidth]{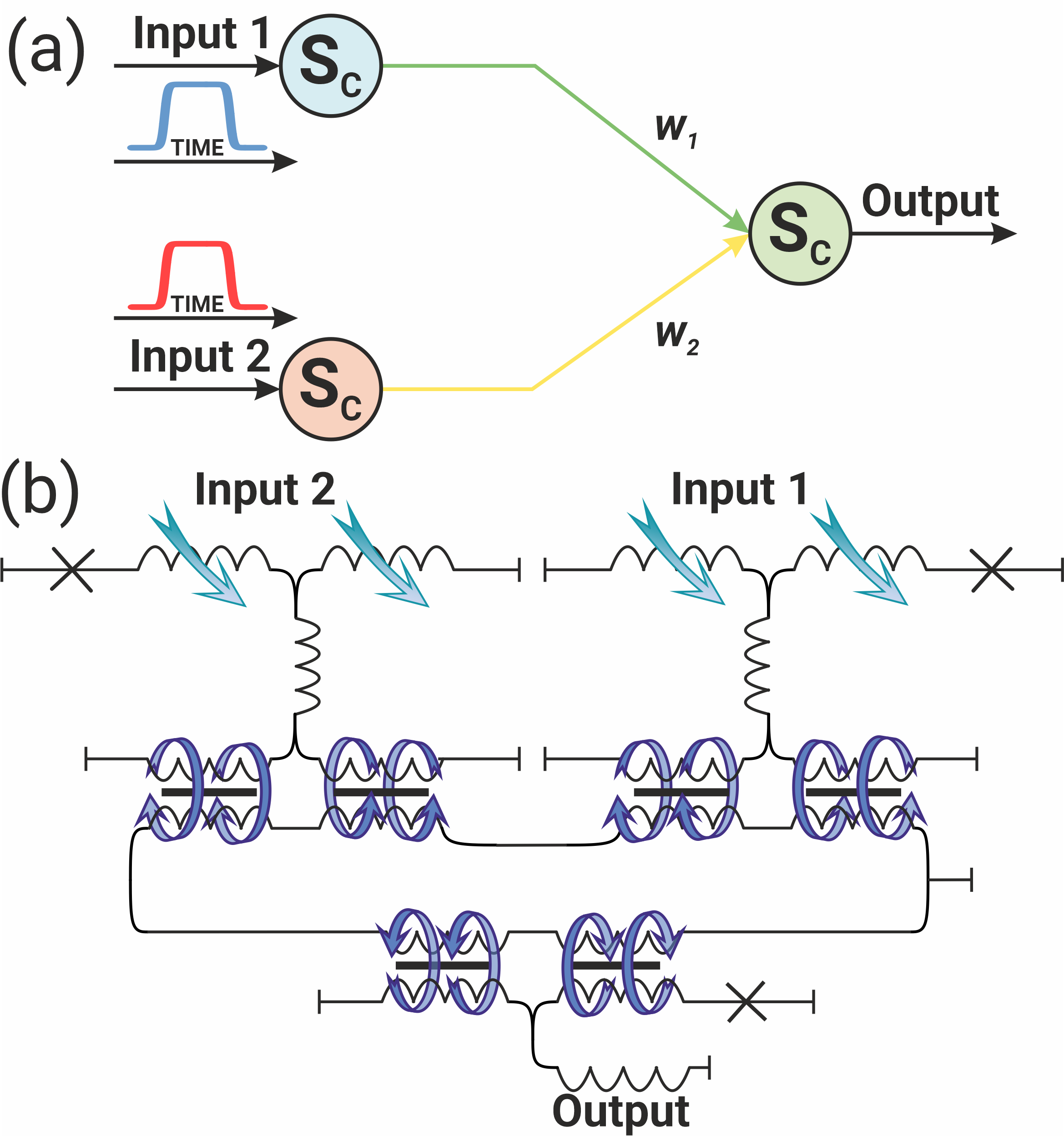}
\caption{
(a) Schematic representation of the 3-neuron XOR/OR network and (b) its superconducting implementation
}
\label{XOR}
\end{figure}

The classical XOR (logical inequality operator) element has two inputs and one output. If the input signals do not match, the output is ``1'', and ``0'' otherwise. The basic neural network implementing XOR consists of three neurons (two input neurons and one output neuron). The inputs of the neural network are supplied with signal in the form of smoothed trapezoid ``1'' or no signal ``0'', see Figure~\ref{XOR}a, respectively. The optimisation problem is reduced to finding such parameters of the system at which the output layer neuron activates according to the XOR truth table. Similar considerations are valid for obtaining a neural network operating according to the OR gate principle.

The discussion of the neural-XOR/OR superconducting circuit based on three adiabatic neurons (shown in Figure~\ref{XOR}b) begins with writing down the corresponding system of equations:

\begin{widetext}
\begin{equation}
 \begin{cases}

(i_{1})_{in1}+(i_{a1})_{in1}+(i_{out1})_{in1}=0, \\

(i_{1})_{in2}+(i_{a1})_{in2}+(i_{out1})_{in2}=0,\\

(\varphi_1)_{in1} - \frac{\varphi_{in1}}{2} + (i_1 l_1)_{in1} = (i_{out1} l_{out1})_{in1}+(i_{s1} l_{s1})_{in1}+(m_2)_{in1} i_{circ}, \\

(\varphi_1)_{in2} - \frac{\varphi_{in2}}{2} +  (i_1 l_1)_{in2} = (i_{out1} l_{out1})_{in2}   + (i_{s1} l_{s1})_{in2}+  (m_2)_{in2} i_{circ},\\

(\varphi_1)_{in1} - \frac{\varphi{in1}}{2} + (i_1 l_1)_{in1}= (i_{out1} l_{out1})_{in1} +(i_{s2} l_{s2})_{out1}-(m_{2}^{*})_{in1} i_{circ},\\

(\varphi_1)_{in2} - \frac{\varphi_{in2}}{2} +  (i_1 l_1)_{in2}= (i_{out1} l_{out1})_{in2}  + (i_{s2} l_{s2})_{in2} - (m_{2}^{*})_{in2} i_{circ},\\

(\varphi_1)_{in1} - \frac{\varphi_{in1}}{2} + (i_1 l_1)_{in1}= (i_{a1} l_{a})_{in1} +\frac{\varphi_{in1}}{2},\\

(\varphi_1)_{in2} - \frac{\varphi_{in2}}{2} + (i_1 l_1)_{in2}= (i_{a1} l_{a})_{in2} +\frac{ \varphi_{in2}}{2},\\

(i_{out1})_{in1}-(i_{s1})_{in1}-(i_{s2})_{in1}=0,\\

(i_{out1})_{in2}-(i_{s1})_{in2}-(i_{s2})_{in2}=0,\\

((l_{t1})_{in1} + (l_{t2})_{in1} + (l_{t1})_{in2} + (l_{t2})_{in2} + (l_{t3})_{out1} + (l_{t4})_{out1}) i_{circ} =\\ (-(i_{s1} m_2)_{in1} - (i_{s1} m_2)_{in2} + (i_2 m_3)_{out1} + (m_2^* i_{s2})_{in1} + (m_2^* i_{s2})_{in2} - (m_3^* i_{a2})_{out1}),\\

(i_2)_{out1} + (i_{a2})_{out1} + (i_{out2})_{out1} = 0,\\
 
(\varphi_2)_{out1} - (m_3)_{out1} i_{circ} + (i_2 l_2)_{out1} = (i_{out2} l_{out2})_{out1},\\

(\varphi_2)_{out1} - (m_3)_{out1} i_{circ} + (i_2 l_2)_{out1} = (i_{a2} l_{a2})_{out1} + (m_3^*)_{out1} i_{circ},

 \end{cases}\label{eq:XOR/OR}
\end{equation}
\end{widetext}
where we preserve the notations according to Figure~\ref{fig6}, but with the subscripts for the neurons in the input ($in1,2$) and output ($out1$) layers (a more detailed scheme with all designations can be found in the Supplementary Materials, see Figure~S1). The input signals defined by expression (\ref{eq:phi_in}) are denoted accordingly $\varphi_{in1,2}$.

\begin{figure}[h!]
\center{\includegraphics[scale=0.35]{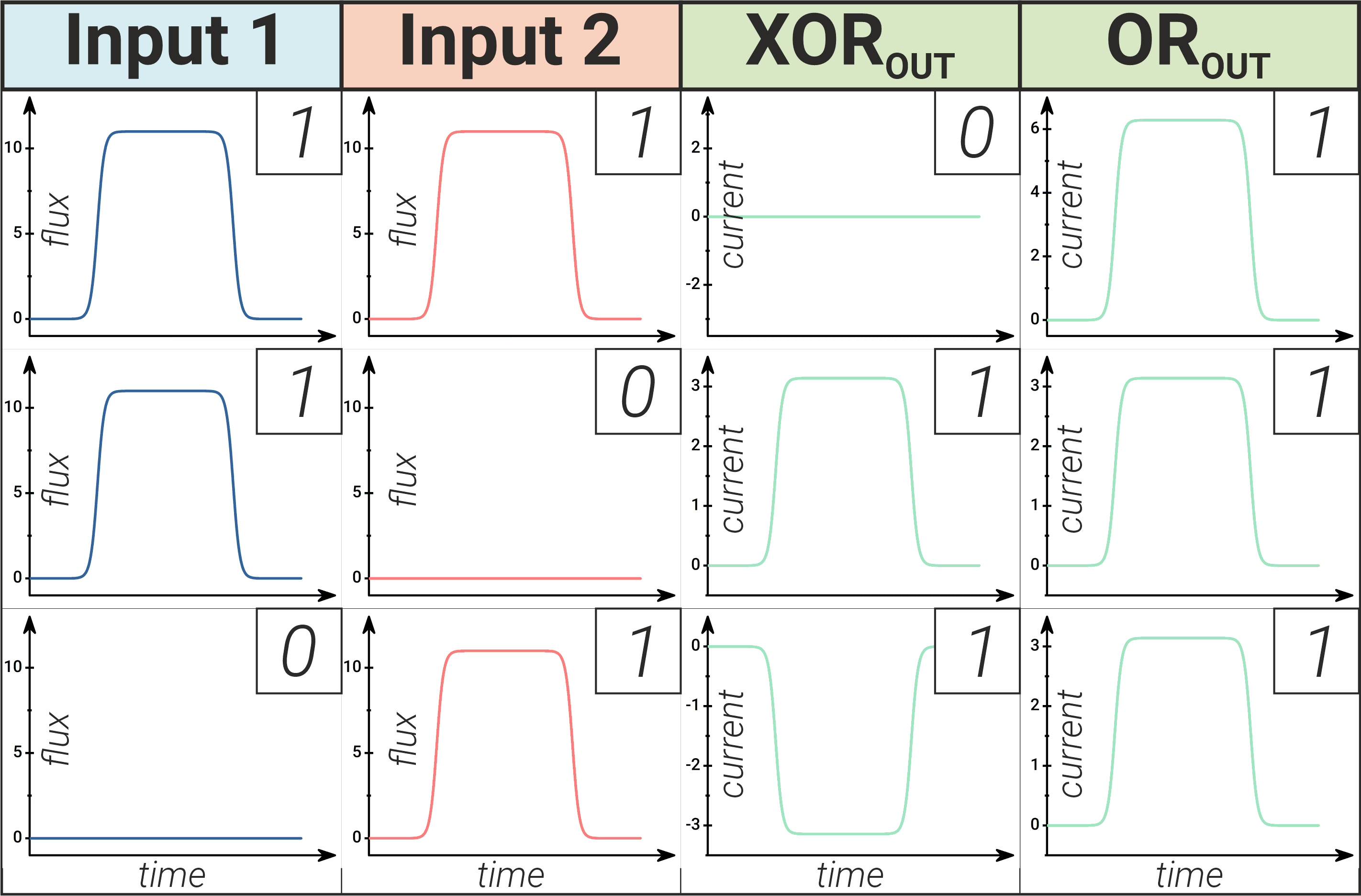}}
\caption{Demonstration of neural network operation as an XOR/OR logic gate.} Synaptic weights are asymmetric/symmetric respectively. The scheme of the neural network is shown in Figure~\ref{XOR}.
\label{fig10}
\end{figure}

Solving the optimisation problem for the system of equations (\ref{eq:XOR/OR}) makes it possible to configure the neural network capable to operate both as an XOR or as an OR logic element, which is quite expected. An obvious choice for such a neural network configuration will be the choice of weight coefficients: they should be asymmetric for XOR, and, on the contrary, they should be symmetric for OR implementation.  
By solving the system of equations (\ref{eq:XOR/OR}) describing the circuit shown in Figure~\ref{XOR}, the truth tables for XOR/OR network implementations were obtained and presented in Figure~\ref{fig10}. The case when there is no signal at the input of both input neurons is not shown: if there is no signal at both inputs of the circuit, there is no signal at the output as well.  

One point is worth mentioning regarding the proposed implementations of the neural networks. Here the XOR output can be of both: positive and negative polarity (see Figure~\ref{fig10}). The OR output with ``1'' in both inputs is twice larger than that with inputs ``1'' + ``0'' or ``0'' + ``1'' (see Figure~\ref{fig10}). This is in contrast to the digital implementations where the output can be ``0'' or ``1'' only.

\section{Conclusion}
In this paper, we demonstrated optimisation algorithm for the parameters of adiabatic neural networks. The algorithm allowed us to find the optimal values for operation of the circuits with different combinations of synapses and neurons, including the ones mimicking logical XOR and OR elements. In addition, a generalisation of this algorithm to neural networks of higher dimensionality consisting of superconducting $S_c$-neurons and synapses was discussed. 

It should be noted that even in the development of such simple neural networks, we faced a significant signal decay problem. For larger neural networks the solution may imply an addition of magnetic flux amplifiers (boosters), well-known in adiabatic superconducting logic \cite{mizushima2023adiabatic}. An utilization of analogue-digital (and, apparently, optical-superconductor) approach for the network implementations is another option.

Regarding the experimental feasibility of the presented schemes, there are a number of experimental works \cite{ionin2023experimentalS, ionin2023experimentalG, he2020compact, takeuchi2023scalable} using a similar technique for the fabrication of Josephson junctions and demonstrating their critical currents in the range of 50 to 150 $\mu$A, corresponding to characteristic values of inductance magnitudes at the level of 2.2 -- 6.6 pH. This confirms the experimental feasibility of the design considerations presented.

\section{ACKNOWLEDGMENTS}
The development of the main concept was carried out with the financial support of the Strategic Academic Leadership Program ``Priority-2030'' (grant from NITU ``MISIS'' No. K2-2022-029). The development of the method of analysing for the evolution of the adiabatic logic cells was carried out with the support of the Grant of the Russian Science Foundation No. 22-72-10075.  A.S. is grateful to the grant 22-1-3-16-1 from the Foundation for the Advancement of Theoretical Physics and Mathematics ``BASIS''. The work of M.B. and I.S. was supported by Rosatom in the framework of the Roadmap for Quantum computing (Contract No. 868-1.3-15/15-2021 dated October 5).

\newpage

\begin{widetext}
    \section{Supplementary Materials}
    \subsection{Analycal expressions of coefficients for ``equations of motion''}

The analytical representation of the coefficients for the ''equations of motion'' of the composite system (2) is given below:

\begin{equation*}
    \begin{split}
      \kappa_{1}^{(2)} & = \kappa_{2}^{(1)} = 
1/(l_{1} l_{a2} l_{o}^2 l_{ps2} m_{1}) (l_{a2} m_{2p} (l_{ao1} l_{ps2} m_{2}+l_{ao1} l_{inp} m_{2p}-l_{r1} m_{2p}) (-l_{k} l_{o} l_{out1}+\\
&l_{ao2} (l_{ps2} m_{2}-l_{ps1} m_{2}^{*}) (-m_{1}^{2} m_{2p}+l_{out1} (2 l_{p} m_{2}+l_{s2} m_{2}+l_{in} m_{2p}+l_{p} m_{2}^{*}))) \\
&(l_{a2} l_{at2} m_{3}+(l_{ao2}-l_{2} l_{at2}) m_{3}^{*})+l_{o} l_{ps2} (m_{1}^{2} m_{2p}-l_{out1} (l_{s2} m_{2}+l_{in} m_{2p}+l_{p} (2 m_{2}+m_{2}^{*}))) \\
&(l_{ao2} m_{3}^{*}+(l_{a2} m_{3}-l_{2} m_{3}^{*}) (l_{out2}+l_{a2} (l_{a2} m_{3}-l_{2} m_{3}^{*})))),\\
\kappa_{1}^{(in)} &=
(l_{w}-(l_{1}-l_{a1}) l_{a2} (-m_{1}^{2} m_{2p} (-l_{ps2} l_{x}+l_{y}+l_{ao2} m_{2p} m_{2}^{*})+l_{out1} (l_{ps1} l_{s2} l_{x} m_{2}+ l_{ps2} l_{y} m_{2} \\
&-l_{is} l_{s2} l_{x} m_{2p}+l_{inp} l_{y} m_{2p}+l_{p} l_{x} (l_{ps1} m_{2}-l_{is} m_{2p})+l_{ao2} m_{2p} (-l_{ps1} m_{2}+ \\
&+ l_{is} m_{2p}) m_{2}^{*})))/(2 l_{1} l_{w}),\\
\kappa_{2}^{(1)} &=
-((l_{a1} l_{q} m_{1} m_{2p} (l_{ps2} m_{2}-l_{ps1} m_{2}^{*}) (l_{a2} m_{3}+l_{out2} m_{3p}))/((l_{r1} (l_{q}-l_{a2} l_{y}) m_{2p}+ \\
&l_{ao1} (l_{ps1} l_{q} m_{2}+l_{a2} l_{ps2} l_{y} m_{2}-l_{is} l_{q} m_{2p}+l_{a2} l_{inp} l_{y} m_{2p})) (-l_{2} l_{at2} l_{f}-l_{a2} l_{out2} l_{ps2} l_{t}+ \\
&l_{a2} l_{out2} m_{2p} m_{2}^{*}+l_{a2} l_{ps2} m_{3}^{2}+l_{out2} l_{ps2} m_{3p}^{2}+l_{2} l_{ps2} m_{3}^{*2})))\\
\kappa_{2}^{(in)} &= 
-(((l_{1}-l_{a1}) l_{q} m_{1} m_{2p} (l_{ps2} m_{2}-l_{ps1} m_{2}^{*}) (l_{a2} m_{3}+l_{out2} m_{3p}))/(2 (l_{r1} (l_{q}-l_{a2} l_{y}) m_{2p}+ \\
&l_{ao1} (l_{ps1} l_{q} m_{2}+l_{a2} l_{ps2} l_{y} m_{2}-l_{is} l_{q} m_{2p}+l_{a2} l_{inp} l_{y} m_{2p})) (-l_{2} l_{at2} l_{f}-l_{a2} l_{out2} l_{ps2} l_{t}+ \\
&l_{a2} l_{out2} m_{2p} m_{2}^{*}+l_{a2} l_{ps2} m_{3}^{2}+l_{out2} l_{ps2} m_{3p}^{2}+l_{2} l_{ps2} m_{3}^{*2}))),\\
\kappa_{in}^{(1)} &=
(l_{a1} m_{1} (l_{2} l_{at2} (-l_{ps} l_{t}+m_{2p}^{2})+l_{a2} l_{out2} (-l_{ps} l_{t}+m_{2p}^{2})+l_{a2} l_{ps} m_{3}^{2}+l_{out2} l_{ps} m_{3p}^{2}+ \\
&l_{2} l_{ps} m_{3}^{*2}))/l_{g},\\
\kappa_{in}^{(1)} &=
(l_{a1} m_{1} (l_{2} l_{at2} (-l_{ps} l_{t}+m_{2p}^{2})+l_{a2} l_{out2} (-l_{ps} l_{t}+m_{2p}^{2})+l_{a2} l_{ps} m_{3}^{2}+l_{out2} l_{ps} m_{3p}^{2}+ \\
&l_{2} l_{ps} m_{3}^{*2}))/l_{g},\\
    \end{split}
\end{equation*}

\begin{equation*}
    \begin{split}
\kappa_{in}^{(2)} & = -((l_{ao1} (l_{ps2} m_{2}-l_{ps1} m_{2}^{*}) (l_{a2} m_{3}+l_{out2} m_{3p}))/l_{g}),\\
\kappa_{in}^{(in)} &= 
-(((l_{1}-l_{a1}) m_{1} (l_{a2} (l_{out2} l_{ps} l_{t}-l_{out2} m_{2p}^{2}-l_{ps} m_{3}^{2})-l_{out2} l_{ps} m_{3p}^{2}+l_{2} (l_{at2} l_{ps} l_{t}- \\
&l_{at2} m_{2p}^{2}-l_{ps} m_{3}^{*2})))/l_{d}),\\
\kappa^{(1)}_{s} &=
(l_{a1} m_{1} (l_{a2} l_{out2} (-l_{s1} l_{t}+l_{s2} l_{t}+m_{2} m_{2p}-m_{2p} m_{2}^{*})+l_{2} l_{at2} (-l_{s1} l_{t}+l_{s2} l_{t}+m_{2}^{2}- \\
&m_{2}^{*2})+l_{a2} (l_{s1}-l_{s2}) m_{3}^{2}+l_{out2} (l_{s1}-l_{s2}) m_{3p}^{2}+l_{2} (l_{s1}-l_{s2}) m_{3}^{*2}))/l_{g},\\
\kappa^{(2)}_{s} &=
(l_{ao1} (2 l_{in}+3 l_{p}+l_{s2}) m_{2}-2 l_{r1} m_{2p}+l_{ao1} (2 l_{in}+3 l_{p}+l_{s1}) m_{2}^{*}) (l_{a2} m_{3}+ l_{out2} m_{3p})/l_{g}\\
\kappa^{(in)}_{s} &=
-(((l_{1}-l_{a1}) m_{1} (l_{2} l_{at2} (l_{s1} l_{t}-l_{s2} l_{t}-m_{2}^{2}+m_{2}^{*2})+l_{a2} l_{out2} (l_{s1} l_{t}-l_{s2} l_{t}-m_{2}^{2}+m_{2}^{*2})+ \\
&l_{a2} (-l_{s1}+l_{s2}) m_{3}^{2}+l_{out2} (-l_{s1}+l_{s2}) m_{3p}^{2}+l_{2} (-l_{s1}+l_{s2}) m_{3}^{*2}))/l_{d}).\\
        \end{split}
\end{equation*}

The following is a list of the notations used in the given coefficients:
$$l_{t} =l_{t1}+l_{t2}+l_{t3}+l_{t4},$$ $$l_{ao1} = l_{1} l_{at1} + l_{a1} l_{out1}, \quad l_{ao2} = l_{2} l_{at2} + l_{a2} l_{out2},$$ $$l_{r1} = (l_{1}+l_{a1}) m_{1}^{2}, \quad l_{ps1} = l_{p}+l_{s1},  \quad l_{ps2} = l_{p}+l_{s2},$$ $$m_{2p} = m_{2}+m_{2}^{*},  \quad m_{3p} = m_{3}+m_{3}^{*},$$
$$l_{ps1} = l_{p}+l_{s1},  \quad l_{ps2} = l_{p}+l_{s2},$$
$$l_{x} = l_{2} l_{at2} l_{t} + l_{a2} l_{out2} l_{t} - l_{a2} m_{3}^{2} - l_{out2} m_{3p}^{2} - l_{2} m_{3}^{*2},$$
\begin{equation*}
    \begin{split}
l_{y} = l_{ao2} m_{2} m_{2p}+&l_{a2} l_{ps1} (-l_{out2} l_{t}+m_{3}^{2})+l_{out2} l_{ps1} m_{3p}^{2}+ \\
&l_{2} l_{ps1} (-l_{at2} l_{t}+m_{3}^{*2})
    \end{split}
\end{equation*}
$$l_{inp} = l_{in}+l_{p}, \quad l_{is} = l_{in}+2 l_{p}+l_{s1},$$
$$ l_{q} = l_{a2} (l_{ps2} l_{x} - l_{ao2} m_{2p} m_{2}^{*}),$$
\begin{equation*}
    \begin{split}
l_{w} = l_{r1} (-l_{q}+&l_{a2} l_{y}) m_{2p}-l_{ao1} (l_{ps1} l_{q} m_{2}+l_{a2} l_{ps2} l_{y} m_{2} \\ 
&-l_{is} l_{q} m_{2p}+l_{a2} l_{inp} l_{y} m_{2p})
    \end{split}
\end{equation*}
$$l_{j} = l_{out12} l_{p}, \quad l_{ps} = 2 l_{p}+l_{s1}+l_{s2},$$
$$l_{k} = 3 l_{p}^{2} + l_{in} l_{ps} + 2 l_{p} l_{s12} + l_{s1} l_{s2},$$
$$l_{b} = 2 l_{j} + l_{a2} l_{out1} l_{ps} + l_{out12} l_{s12},$$
$$ l_{s12} = l_{s1}+l_{s2}, \quad l_{at1} = l_{a1}+l_{out1}, \quad l_{at2} = l_{a2}+l_{out2},$$
$$l_{out12} = l_{out1} l_{out2}, \quad l_{ps12} = l_{ps1} + l_{ps2},$$

\begin{equation*}
    \begin{split}
 l_{u} =& l_{a2} (2 l_{j} (-l_{s12} l_{t} +
        m_{2}^{2}+m_{2} m_{2}^{*}+m_{2}^{*2})+l_{out12} (-3 l_{p}^{2} l_{t}-l_{in} l_{ps} l_{t}-l_{s1} l_{s2} l_{t}+l_{s2} m_{2}^{2}+ l_{in} m_{2p}^{2}+l_{s1} m_{2}^{*2})+l_{k} l_{out1} m_{3}^{2} + \\
        &m_{1}^{2} (l_{out2} l_{ps} l_{t}-l_{out2} m_{2p}^{2}-l_{ps} m_{3}^{2}))+l_{out2} (l_{k} l_{out1}- l_{ps} m_{1}^{2}) m_{3p}^{2}+l_{2} (-l_{b} l_{in} l_{t}-l_{t} (3 l_{out12} l_{p}^{2}+ \\
        &2 l_{j} l_{s12}+l_{out12} l_{s1} l_{s2}-l_{out2} l_{ps} m_{1}^{2})+(2 l_{j}+ l_{out12} l_{s2}- l_{out2} m_{1}^{2}) m_{2}^{2}+l_{in} l_{out12} m_{2p}^{2}+m_{2}^{*} (2 l_{j} m_{2p}+ \\
        &l_{out12} l_{s1} m_{2}^{*}-l_{out2} m_{1}^{2} (2 m_{2}+m_{2}^{*}))+ l_{a2} (m_{1}^{2} (l_{ps} l_{t} - m_{2p}^{2})+l_{out1} (-3 l_{p}^{2} l_{t}-l_{s1} l_{s2} l_{t}+l_{s2} m_{2}^{2}+ \\
        &l_{in} m_{2p}^{2}+l_{s1} m_{2}^{*2}+2 l_{p} (-l_{s12} l_{t}+ m_{2}^{2}+ m_{2} m_{2}^{*}+ m_{2}^{*2})))+(l_{k} l_{out1}-l_{ps} m_{1}^{2}) m_{3}^{*2}), \\
    l_{e} =&  l_{out2} m_{1}^{2} (2 l_{p} l_{t}+l_{s12} l_{t}-m_{2p}^{2})+2 l_{j} (-l_{s12} l_{t}+m_{2}^{2}+m_{2} m_{2}^{*}+m_{2}^{*2})+l_{out12} (-3 l_{p}^{2} l_{t}+ l_{s2} m_{2}^{2} + l_{s1} (-l_{s2} l_{t}+m_{2}^{*2}))+\\
        &l_{a1} l_{out2} (-3 l_{p}^{2} l_{t}+l_{s2} m_{2}^{2}+l_{s1} (-l_{s2} l_{t}+m_{2}^{*2})+2 l_{p} (-l_{s12} l_{t}+ m_{2}^{2}+m_{2} m_{2}^{*} + m_{2}^{*2}))+(l_{at1} (3 l_{p}^{2}+2 l_{p} l_{s12}+l_{s1} l_{s2}) -\\
        &l_{ps} m_{1}^{2}) m_{3}^{2}+l_{in} (-2 l_{j} l_{t}-l_{out12} l_{s12} l_{t}+ l_{out12} m_{2p}^{2}+ l_{a1} l_{out2} (-l_{ps} l_{t}+m_{2p}^{2})+l_{at1} l_{ps} m_{3}^{2}), \\
l_{h} =& -l_{b} l_{in} l_{t}-l_{a2} l_{out1} (3 l_{p}^{2}+2 l_{p} l_{s12}+l_{s1} l_{s2}) l_{t}+l_{t} (3 l_{out12} l_{p}^{2}+2 l_{j} l_{s12}+l_{out12} l_{s1} l_{s2}- l_{at2} l_{ps} m_{1}^{2})+(2 l_{j}+l_{out12} l_{s2}+ \\ 
&l_{a2} l_{out1} (2 l_{p}+l_{s2})-l_{at2} m_{1}^{2}) m_{2}^{2}+ l_{in} (l_{a2} l_{out1}+l_{out12}) m_{2p}^{2}- 2 (l_{j}+l_{a2} l_{out1} l_{p}-l_{at2} m_{1}^{2}) m_{2} m_{2}^{*}+(2 l_{j}+l_{out12} l_{s1}+ \\
&l_{a2} l_{out1} (2 l_{p}+l_{s1})-    l_{at2} m_{1}^{2}) m_{2}^{*2}+ l_{a1} l_{at2} (-3 l_{p}^{2} l_{t}+l_{in} l_{ps} l_{t}-l_{s1} l_{s2} l_{t}+l_{s2} m_{2}^{2}-l_{in} m_{2p}^{2}+l_{s1} m_{2}^{*2}+2 l_{p} (-l_{s12} l_{t}+m_{2}^{2}+\\
        &m_{2} m_{2}^{*}+ m_{2}^{*2}))+l_{a1} l_{k} m_{3}^{*2}+(l_{k} l_{out1}-l_{ps} m_{1}^{2}) m_{3}^{*2}\\
l_{g} =& l_{a1} l_{u}+l_{1} (l_{a2} l_{e}+l_{2} l_{h}+(2 l_{in} l_{j}+ 3 (l_{out12}+l_{a1} l_{out2}) l_{p}^{2}+l_{a1} l_{in} l_{out2} l_{ps}+2 l_{j} l_{s12}+l_{in} l_{out12} l_{s12}+2 l_{a1} l_{out2} l_{p} l_{s12}+\\
&l_{out12} l_{s1} l_{s2}+ l_{a1} l_{out2} l_{s1} l_{s2}-l_{out2} l_{ps} m_{1}^{2}) m_{3}^{2}+2 l_{out2} (l_{at1} l_{k}-l_{ps} m_{1}^{2}) m_{3} m_{3}^{*}+l_{out2} (l_{at1} l_{k}-l_{ps} m_{1}^{2}) m_{3}^{*2}),\\
l_{d} = &2(l_{a1} l_{u}+l_{1} (l_{2} l_{h}+(2 l_{in} l_{j}+3 (l_{out12}+l_{a1} l_{out2}) l_{p}^{2}+2 l_{j} l_{s12}+l_{in} l_{out12} l_{s12}+l_{out12} l_{s1} l_{s2}+l_{a1} l_{out2} (l_{in} l_{ps}+2 l_{p} l_{s12}+ \\
&l_{s1} l_{s2})-l_{out2} l_{ps} m_{1}^{2}) m_{3}^{2}-l_{a2} (l_{in} (2 l_{j}+l_{a1} l_{out2} l_{ps}+l_{out12} l_{s12}) l_{t}+2 l_{j} (l_{s12} l_{t}-m_{2}^{2})+l_{out2} m_{1}^{2} (-l_{ps} l_{t}+m_{2}^{2})-\\
&l_{in} (l_{out12}+l_{a1} l_{out2}) m_{2p}^{2}+l_{out12} (3 l_{p}^{2} l_{t}+l_{s1} l_{s2} l_{t}-l_{s2} m_{2}^{2}-l_{s1} m_{2}^{*2})+l_{a1} l_{out2} (3 l_{p}^{2} l_{t}+2 l_{p} l_{s12} l_{t}+l_{s1} l_{s2} l_{t}-l_{s2} m_{2}^{2}- \\
&2 l_{p} m_{2} m_{2p}-(2 l_{p}+l_{s1}) m_{2}^{*2})-m_{2}^{*} (2 l_{j} (m_{2}-m_{2}^{*})+l_{out2} m_{1}^{2} (2 m_{2}+m_{2}^{*}))-l_{at1} l_{k} m_{3}^{2}+l_{ps} m_{1}^{2} m_{3}^{2})+2 l_{out2} (l_{at1} l_{k}-\\
&l_{ps} m_{1}^{2}) m_{3} m_{3}^{*}+l_{out2} (l_{at1} l_{k}-l_{ps} m_{1}^{2}) m_{3}^{*2})), \\
l_{f} =& l_{ps2} l_{t} - m_{2p} m_{2}^{*}, l_{o} =  l_{2} l_{at2} l_{f}+l_{a2} l_{out2} l_{ps2} l_{t}-l_{a2} l_{out2} m_{2p} m_{2}^{*}-l_{a2} l_{ps2} m_{3}^{2}-l_{out2} l_{ps2} m_{3p}^{2}-l_{2} l_{ps2} m_{3}^{*2}
    \end{split}
\end{equation*}
\end{widetext}

\subsection{Hamiltonian formalism for two coupled neurons}

The study of the dynamic properties of connected neurons can also be carried out within the framework of the Hamiltonian formalism. Herewith our system can be represented as two bound particles with momenta $p_{n}=c_{n} \dot{\varphi_{n}}$, and therefore the motion of particles obeys the classical Hamilton's equations:
\begin{equation}
    \begin{cases}
        \dot{\varphi_{n}} = \frac{\partial H}{\partial p_{n}}, \\
        \dot{p_{n}} = - \frac{\partial H}{\partial \varphi_{n}},
    \end{cases}
    \label{EQ:dH}
\end{equation}
where $H$ is a Hamiltonian (an integral of motion).

By integrating the system (\ref{EQ:dH}) we can find the momentum parts of the Hamiltonian, which have the standard form $p^2_{n} / (2 c_{n})$. Also from Eq.~(5) it can be seen that $\dot{p_{n}}=i_{n}-i_{C_{n}}\sin\varphi_{n}$, therefore using the Hamiltonian's feature $\kappa \equiv \kappa_{1}^{(2)} = \kappa_{2}^{(1)}$ one gets the form:

\begin{equation}
\begin{aligned}
        H = \sum_{n=1,2}\left( H_{\omega_{n}} - i_{C_{n}}\cos\varphi_{n}\right) + \kappa\varphi_{1}\varphi_{2} - \\
        \left(\frac{\left(\kappa_{1}^{(in)}\right)^2}{2 \kappa_{1}^{(1)}} + \frac{\left(\kappa_{2}^{(in)}\right)^2}{2 \kappa_{2}^{(2)}} \right) \varphi^{2}_{in}.
    \label{EQ:Hphi1}
\end{aligned}
\end{equation}
Here the first term defines the Hamiltonians of individual neurons
\begin{equation*}
 H_{\omega_{n}} = \frac{p^2_{n}}{2 c_{n}} + \frac{\omega^2_{n}}{2}\left(\varphi_{n} + \frac{\kappa_{n}^{(in)}}{\kappa_{n}^{(n)}}\varphi_{in}\right)^2,
 \omega^2_{n} = \kappa_{n}^{(n)}\;(n=1,2).   
\end{equation*}
The second term in Eq.~(\ref{EQ:Hphi1}) defines the interaction between neurons, and the last term is responsible for the dynamic control of neurons by the external magnetic flux $\varphi_{in}$ according to (1).

Using the example of two interacting neurons within the framework of the Hamiltonian formalism, it is clearly seen that the dynamic processes in the system resemble two interacting nonlinear oscillators, the coupling strength of which is linear relative to the phases at each of the Josephson contacts. Using this approach, the numerical analysis of the system is reduced to solving coupled differential equations of the first order (in contrast to (5), where it is necessary to solve a system of differential equations of the second order), which greatly simplifies further consideration of more complex systems of coupled neurons and synapses.

\subsection{Superconducting XOR/OR network scheme with notations}
\begin{figure}
    \centering
    \includegraphics[width=0.98\linewidth]{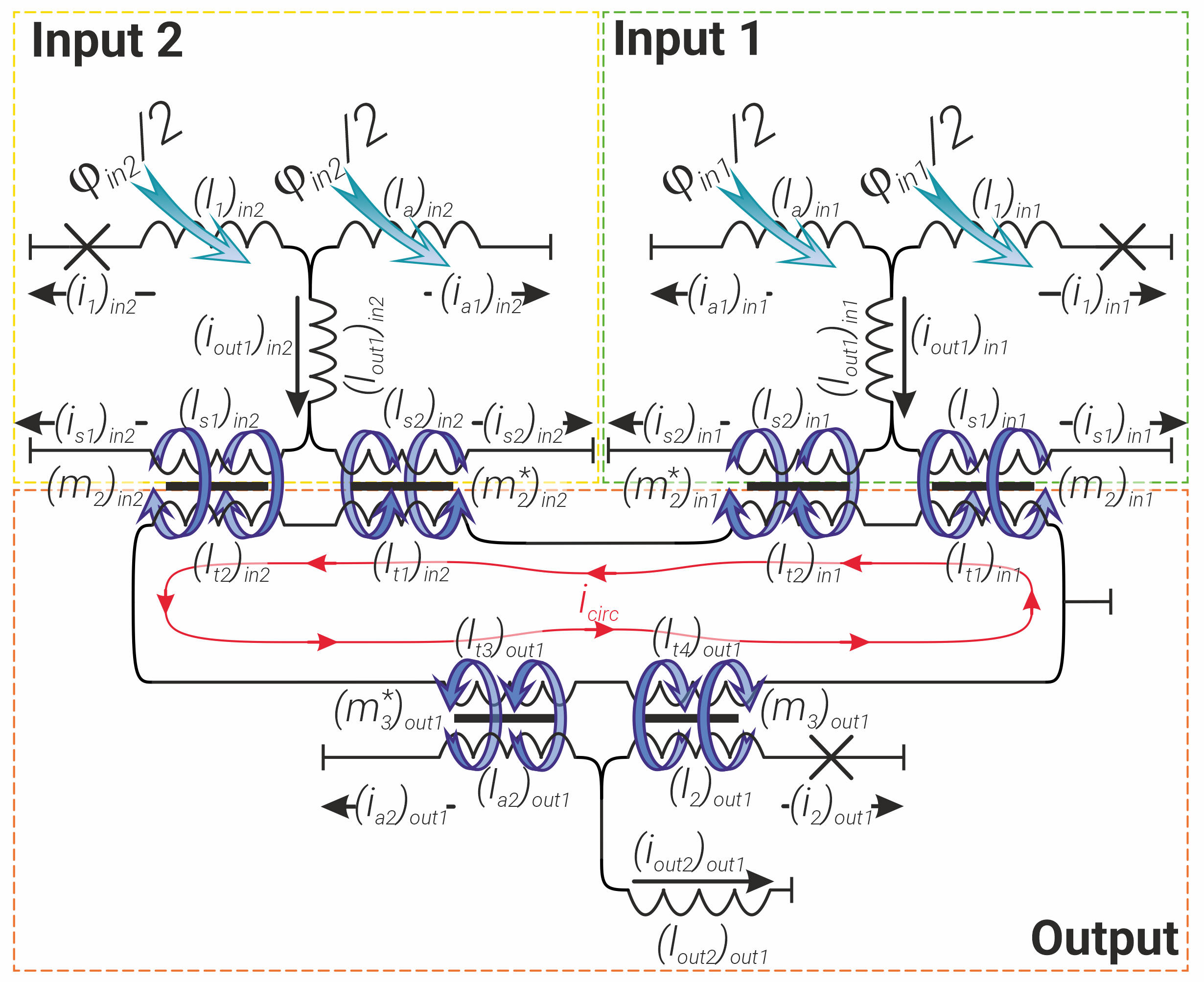}
    \caption{Schematic representation of the 3-neuron XOR/OR network in its superconducting implementation}
    \label{fig:XOR 3N}
\end{figure}

Here is a more detailed scheme of the XOR/OR network with all the notations used in the system of equations (9) -- Figure~\ref{fig:XOR 3N}.

\bibliographystyle{apsrev4-1}
\bibliography{bibliography}

\end{document}